\documentclass[12pt,letterpaper]{article}
\usepackage{lmodern}
\usepackage{amssymb,amsmath}
\usepackage{ifxetex,ifluatex}
\usepackage{fixltx2e} % provides \textsubscript
\ifnum 0\ifxetex 1\fi\ifluatex 1\fi=0 % if pdftex
  \usepackage[T1]{fontenc}
  \usepackage[utf8]{inputenc}
\else % if luatex or xelatex
  \ifxetex
    \usepackage{mathspec}
  \else
    \usepackage{fontspec}
  \fi
  \defaultfontfeatures{Ligatures=TeX,Scale=MatchLowercase}
\fi
\IfFileExists{upquote.sty}{\usepackage{upquote}}{}
\IfFileExists{microtype.sty}{%
\usepackage{microtype}
\UseMicrotypeSet[protrusion]{basicmath} % disable protrusion for tt fonts
}{}
\usepackage[margin=1in]{geometry}
\usepackage{hyperref}
\hypersetup{unicode=true,
            pdftitle={Visualizing Tests for Equality of Covariance Matrices},
            pdfauthor={Michael Friendly and Matthew Sigal},
            pdfkeywords={Box M test; HE plots; MANOVA; graphics},
            pdfborder={0 0 0},
            breaklinks=true}
\usepackage{graphicx,grffile}
\makeatletter
\def\maxwidth{\ifdim\Gin@nat@width>\linewidth\linewidth\else\Gin@nat@width\fi}
\def\maxheight{\ifdim\Gin@nat@height>\textheight\textheight\else\Gin@nat@height\fi}
\makeatother
\setkeys{Gin}{width=\maxwidth,height=\maxheight,keepaspectratio}
\IfFileExists{parskip.sty}{%
\usepackage{parskip}
}{% else
\setlength{\parindent}{0pt}
\setlength{\parskip}{6pt plus 2pt minus 1pt}
}
\ifx\paragraph\undefined\else
\let\oldparagraph\paragraph
\renewcommand{\paragraph}[1]{\oldparagraph{#1}\mbox{}}
\fi
\ifx\subparagraph\undefined\else
\let\oldsubparagraph\subparagraph
\renewcommand{\subparagraph}[1]{\oldsubparagraph{#1}\mbox{}}
\fi
\usepackage{bm}
\usepackage{xspace}
\usepackage{amssymb}  % for \mathbb{}

\newcommand*{\figref}[1]{Figure~\ref{#1}}
\newcommand*{\tabref}[1]{Table~\ref{#1}}
\newcommand*{\secref}[1]{Section~\ref{#1}}
\renewcommand*{\eqref}[1]{Eqn.~(\ref{#1})}

\newcommand{\mat}[1]{\ensuremath{\bm{#1}}}               % matrix (bold)
\renewcommand{\vec}[1]{\ensuremath{\bm{#1}}}
\newcommand{\trans}{\ensuremath{^\mathsf{T}}}            % transpose
\newcommand*{\comma}{\: ,}%                      punct after displaymath [made smaller]
\newcommand*{\period}{\: .}

\renewcommand*{\det}[1]{\ensuremath{\mathrm{det} (#1)}}

\newcommand*{\trace}[1]{\ensuremath{\mathrm{tr} (\mat{#1})}}
\newcommand*{\dev}[1]{(#1 - \bar{#1})}
\newcommand*{\inv}[1]{\ensuremath{\mat{#1}^{-1}}}

\newcommand*{\Real}[1]{\mathbb{R}^{#1}}     % nicer looking than \Re
\newcommand*\widebar[1]{\overline{#1}}
\let\proglang=\textsf
\newcommand{\R}{\proglang{R}\xspace}
\newcommand{\code}[1]{\texttt{#1}}
\newcommand{\func}[1]{\code{#1()\xspace}}
\newcommand{\pkg}[1]{\textsf{#1}}

\newcommand{\class}[1]{\texttt{"#1"}}

\usepackage{setspace}
 \def\spacingset#1{\renewcommand{\baselinestretch}%
 {#1}\small\normalsize}
 \spacingset{1.1}

\let\rmarkdownfootnote\footnote%
\def\footnote{\protect\rmarkdownfootnote}

\usepackage{titling}

  \title{Visualizing Tests for Equality of Covariance Matrices}
  \pretitle{\vspace{\droptitle}\centering\huge}
  \posttitle{\par}
  \author{Michael Friendly and Matthew Sigal}
  \preauthor{\centering\large\emph}
  \postauthor{\par}
  \date{}
  \predate{}\postdate{}

\begin{document}
\maketitle
\begin{abstract}
This paper explores a variety of topics related to the question of
testing the equality of covariance matrices in multivariate linear
models, particularly in the MANOVA setting. The main focus is on
graphical methods that can be used to address the evaluation of this
assumption. We introduce some extensions of data ellipsoids,
hypothesis-error (HE) plots and canonical discriminant plots and
demonstrate how they can be applied to the testing of equality of
covariance matrices. Further, a simple plot of the components of Box's M
test is proposed that shows \emph{how} groups differ in covariance and
also suggests other visualizations and alternative test statistics.
These methods are implemented and freely available in the
\textbf{heplots} and \textbf{candisc} packages for R. Examples from the
paper are available in supplementary materials.
\end{abstract}
\providecommand{\keywords}[1]{\textbf{\textit{Keywords---}} #1}

\newpage

\spacingset{1.1}

\section{Introduction}\label{introduction}

\begin{quote}
\emph{To make the preliminary test on variances is rather like putting
to sea in a rowing boat to find out whether conditions are sufficiently
calm for an ocean liner to leave port.} --- G. E. P. Box (1953)
\end{quote}

This paper concerns the extension of tests of homogeneity of variance
from the classical univariate ANOVA setting to the analogous
multivariate (MANOVA) setting. Such tests are a routine but important
aspect of data analysis, as particular violations can drastically impact
model estimates (Lix \& Keselman, 1996). In the multivariate context,
the following questions and topics are of main interest here:

\begin{itemize}
\item
  \textbf{Visualization}: How can we visualize differences among group
  variances and covariance matrices, perhaps in a way that is analogous
  to what is done to visualize differences among group means?
  Multivariate linear models (MLMs) present additional challenges for
  data visualization because we often want to see the effects for a
  collection of response variables \emph{simultaneously}, which pushes
  the boundaries of typical graphical displays. As will be illustrated,
  differences among covariance matrices can be comprised of spread in
  overall size (``scatter'') and shape (``orientation''). When there are
  more than a few response variables, what low-dimensional views can
  show the most interesting properties related to the equality of
  covariance matrices?
\item
  \textbf{Other test statistics}: Test statistics for MANOVA and for
  equality of covariance matrices are based on properties of eigenvalues
  of various matrices. Available tests statistics for mean differences
  suggest alternatives for the question of equality of covariance
  matrices.
\end{itemize}

The following subsections provide a capsule summary of the issues in
this topic. Most of the discussion is couched in terms of a one-way
design for simplicity, but the same ideas can apply to two-way (and
higher) designs, where a ``group'' factor is defined as the product
combination (interaction) of two or more factor variables. When there
are also numeric covariates, this topic can also be extended to the
multivariate analysis of covaraiance (MANCOVA) setting. This can be
accomplished by applying these techniques to the residuals from
predictions by the covariates alone.

\subsection{Homogeneity of Variance in Univariate
ANOVA}\label{homogeneity-of-variance-in-univariate-anova}

In classical (Gaussian) univariate ANOVA models, the main interest is
typically on tests of mean differences in a response \(y\) according to
one or more factors. The validity of the typical \(F\) test, however,
relies on the assumption of \emph{homogeneity of variance}: all groups
have the same (or similar) variance, \[
\sigma_1^2 = \sigma_2^2 = \cdots = \sigma_g^2 \period
\]

It turns out that the \(F\) test for differences in means is relatively
robust to violation of this assumption (Harwell, Rubinstein, Hayes, \&
Olds, 1992), as long as the group sizes are roughly equal.\footnote{If
  group sizes are greatly unequal \textbf{and} homogeneity of variance
  is violated, then the \(F\) statistic is too liberal (\(p\) values too
  large) when large sample variances are associated with small group
  sizes. Conversely, the \(F\) statistic is too conservative if large
  variances are associated with large group sizes.}

A variety of classical test statistics for homogeneity of variance are
available, including Hartley's \(F_{max}\) (Hartley, 1950), Cochran's
\emph{C} (Cochran, 1941),and Bartlett's test (Bartlett, 1937), but these
have been found to have terrible statistical properties (Rogan \&
Keselman, 1977), which prompted Box's famous quote.

Levene (1960) introduced a different form of test, based on the simple
idea that when variances are equal across groups, the average absolute
values of differences between the observations and group means will also
be equal, i.e., substituting an \(L_1\) norm for the \(L_2\) norm of
variance. In a one-way design, this is equivalent to a test of group
differences in the means of the auxilliary variable
\(z_{ij} = \vert y_{ij} - \bar{y}_i \vert\).

More robust versions of this test were proposed by Brown \& Forsythe
(1974). These tests substitute the group mean by either the group median
or a trimmed mean in the ANOVA of the absolute deviations, and should be
almost always preferred to Levene's version (which unfortunately was
adopted as the default in some software such as SPSS). See Conover,
Johnson, \& Johnson (1981) for an early review and Gastwirth, Gel, \&
Miao (2009) for a general discussion of these tests. In what follows, we
refer to this class of tests as ``Levene-type'' tests and suggest a
multivariate extension described in the supplementary appendix
(\secref{sec:supmat}).

\subsection{Homogeneity of variance in
MANOVA}\label{homogeneity-of-variance-in-manova}

MANOVA focuses on testing differences among mean vectors, \[
H_0 : \vec{\mu}_1 = \vec{\mu}_2 = \cdots = \vec{\mu}_g \period
\] However, the standard test statistics (Wilks' Lambda,
Hotelling-Lawley trace, Pillai-Bartlett trace, Roy's maximum root) rely
upon the analogous assumption that the within-group covariance matrices
for all groups are equal, \[
\mathbf{\Sigma}_1 = \mathbf{\Sigma}_2 = \cdots = \mathbf{\Sigma}_g \period
\]

In the multivariate setting, there has been considerable attention to
the sensitivity of these tests to both non-normality and lack of
equality of covariance matrices, largely through simulation studies
(e.g., Finch \& French, 2013; Hakstian, Roed, \& Lind, 1979). Most of
these have been conducted in the simple case of two-group designs (where
Hotelling's \(T^2\) is the equivalent of all the standard tests) or in
one-way designs. A classic study in this area is Olson (1974), that
recommended:

\begin{quote}
for protection against nonnormality and heterogeneity of covariance
matrices, the largest-root test should be avoided, while the
Pillai-Bartlett trace test may be recommended as the most robust of the
MANOVA tests, with adequate power to detect true differences in a
variety of situations (p.~894).
\end{quote}

We mention in passing that, with a burgeoning interesting in robust
methods over the last few decades, there have been a variety of
proposals for how to conduct robust tests of differences on mean
vectors, mostly in the one-way MANOVA setting (e.g., Aelst \& Willems,
2011; Todorov \& Filzmoser, 2010). Generally speaking, these involve
using more robust alternatives for mean vectors (medians, trimmed means,
rank-based methods) and for covariance matrices (e.g., minimum
covariance determinant (MCD) and minimum volume ellipsoid (MVE)).

Yet, there has not been as much attention paid to the second-order
problem of assessing equality of covariance matrices. Box's M test,
described below, remains the main procedure readily available in
statistical software for this problem. The properties and alternatives
to Box's test have not been widely studied (some exceptions are O'Brien,
1992; Tiku \& Balakrishnan, 1984).

However, beyond issues of robustness, the question of equality of
covariance matrices is often of general interest itself. For instance,
variability is often an important issue in studies of strict equivalence
in laboratories comparing across multiple patient measurements and in
other applied contexts (see Gastwirth et al., 2009 for other exemplars).
Moreover the outcome of such tests often have important consequences for
the details of a main method of analysis. Just as the Welsh \(t\)-test
(Welch, 1947) is now commonly used and reported for a two-group test of
differences in means under unequal variances, a preliminary test of
equality of covariance matrices is often used in discriminant analysis
to decide whether linear (LDA) or quadratic discriminant analysis (QDA)
should be applied in a given problem. In such cases, the data at hand
should inform the choice of statistical analysis to utilize.

\subsection{Assessing heterogeneity of covariance matrices: Box's M
test}\label{assessing-heterogeneity-of-covariance-matrices-boxs-m-test}

Box (1949) proposed the following likelihood-ratio test (LRT) statistic
for testing the hypothesis of equal covariance matrices,

\begin{equation}\label{eq:boxm}
M = (N -g) \ln \vert \mathbf{S}_p \vert - \sum_{i=1}^g (n_i -1) \ln \vert \mathbf{S}_i \vert \comma
\end{equation}

where \(N = \sum n_i\) is the total sample size and
\(\mathbf{S}_p = (N-g)^{-1} \sum_{i=1}^g (n_i - 1) \mathbf{S}_i\) is the
pooled covariance matrix. \(M\) can thus be thought of as a ratio of the
determinant of the pooled \(\mathbf{S}_p\) to the geometric mean of the
determinants of the separate \(\mathbf{S}_i\).

In practice, there are various transformations of the value of \(M\) to
yield a test statistic with an approximately known distribution (Timm,
1975). Roughly speaking, when each \(n_i > 20\), a \(\chi^2\)
approximation is often used; otherwise an \(F\) approximation is known
to be more accurate.

Asymptotically, \(-2 \ln (M)\) has a \(\chi^2\) distribution. The
\(\chi^2\) approximation due to Box (1949, 1950) is that \[
X^2 = -2 (1-c_1) \ln (M) \quad \sim \quad \chi^2_{df}
\] with \(df = (g-1) p (p+1)/2\) degrees of freedom, and a bias
correction constant: \[
c_1 = \left( 
\sum_i \frac{1}{n_i -1}
- \frac{1}{N-g}
\right)
\frac{2p^2 +3p -1}{6 (p+1)(g-1)} \period
\]

In this form, Bartlett's test for equality of variances in the
univariate case is the special case when there is only one response
variable, so Bartlett's test is sometimes used as univariate follow-up
to determine which response variables show heterogeneity of variance.

Yet, like its univariate counterpart, Box's test is well-known to be
highly sensitive to violation of (multivariate) normality and the
presence of outliers. For example, Tiku \& Balakrishnan (1984) concluded
from simulation studies that the normal-theory LRT provides poor control
of Type I error under even modest departures from normality. O'Brien
(1992) proposed some robust alternatives, and showed that Box's normal
theory approximation suffered both in controlling the null size of the
test and in power. Zhang \& Boos (1992) also carried out simulation
studies with similar conclusions and used bootstrap methods to obtain
corrected critical values.

\subsection{Visualizing heterogeneity}\label{visualizing-heterogeneity}

The goal of this article is to use the above background as a platform
for discussing approaches to visualizing and testing the heterogeneity
of covariance matrices in multivariate designs. While researchers often
rely on a single number to determine if their data have met a particular
threshold, such compression will often obscure interesting information,
particularly when a test concludes that differences exist, and one is
left to wonder ``why?''. It is within this context where, again,
visualizations often reign supreme. In fact, we find it somewhat
surprising that this issue has not been addressed before graphically in
any systematic way.

In this article, we propose three visualization-based approaches to
questions of heterogeneity of covariance in MANOVA designs: (a) direct
visualization of the information in the \(\mat{S}_i\) and \(\mat{S}_p\)
using \emph{data ellipsoids} to show size and shape as minimal schematic
summaries; (b) a simple dotplot of the components of Box's M test: the
log determinants of the \(\mat{S}_i\) together with that of the pooled
\(\mat{S}_p\). Extensions of these simple plots raise the question of
whether measures of heterogeneity other than that captured in Box's test
might also be useful; and, (c) the connection between Levene-type tests
and an ANOVA (of centered absolute differences) suggests a parallel with
a multivariate extension of Levene-type tests and a MANOVA. We explore
this with a version of Hypothesis-Error (HE) plots we have found useful
for visualizing mean differences in MANOVA designs.

Accordingly, the following sections introduce and apply our conceptual
framework for general graphical methods for visualizing data in relation
to MLM-related questions and their applications. This is based on the
simple ideas that: (a) a \emph{data ellipsoid} provides a visual summary
of location and scatter of a multivariate sample; (b) these can be
combined in various ways to give visual tests of group differences in
means and covariance matrices; and, (c) when there are more than just a
few response variables, a reduced-rank (canonical) transformation
provides an appealing way to visualize these effects in an optimal
low-dimensional approximation.

\secref{sec:vis-boxm} introduces some novel visualizations of the
components related to Box's test, which in turn suggest other possible
test statistics that deserve further study. Section 2 in the
Supplementary Appendix describes a multivariate generalization of
Levene's test within the HE plot framework that yields attractive and
useful displays.

A different graphical approach to the main question is to consider
multivariate dispersion in terms of distances of the points from their
centroids; this is illustrated in the Supplementary Materials. These
methods are all implemented in R (R Core Team, 2015), principally in the
\textsf{heplots} and \textsf{candisc} packages.\footnote{The complete
  \R code for our examples is provided in the Supplementary Materials,
  and are hosted online at \url{https://mattsigal.github.io/eqcov_supp/}}.

\section{Visualizing covariance matrices}\label{sec:vis-cov}

Before diving into details and statistical tests, it is useful to see
how to visualize covariance matrices themselves. We do this using the
graphical analog of minimally sufficient statistics
\((\bar{\vec{y}}_i, \mat{S}_i)\) for the MANOVA problem--- a
\emph{minimally sufficient graphical display}. This graphical principle
has been called \emph{visual thinning} (Friendly, 2007): reducing a
graphical display to the essentials of what you want to see by relying
upon statistics that most efficiently capture the parameters of
interest. In multivariate displays, this usually means replacing data
points by well-chosen visual summaries.

\subsection{Data ellipsoids}\label{data-ellipsoids}

The essential idea (Dempster, 1969; Friendly, Monette, \& Fox, 2013) is
that for a \(p\)-dimensional sample, \(\mat{Y}_{n \times p}\), the
\(p \times p\) covariance matrix \(\mat{S}\) can be represented by the
\(p\)-dimensional \emph{concentration} or \emph{data ellipsoid},
\(\mathcal{E}_c\) of size (``radius'') \(c\). This is defined as the set
of all points \(\vec{y}\) satisfying

\begin{equation}\label{eq:dsq}
\mathcal{E}_c ( \widebar{\vec{y}},  \mat{S} )
:= \{ \vec{y} :
\dev{\vec{y}}\trans \, \inv{S} \, \dev{\vec{y}} \le c^2 \} \period
\end{equation}

It is readily seen that the quadratic form in \eqref{eq:dsq} corresponds
to the set of points whose squared Mahalanobis distances
\(D^2_M (\vec{y}) = \dev{\vec{y}}\trans \, \inv{S} \, \dev{\vec{y}}\),
from the centroid of the sample,
\(\bar{\vec{y}} = (\bar{y}_1, \bar{y}_2, \dots , \bar{y}_p)\trans\), are
less than or equal to \(c^2\).

When the variables are multivariate normal, the data ellipsoid
approximates a contour of constant density in their joint distribution.
In this case \(D^2_M (\vec{y})\) has a large-sample \(\chi^2_p\)
distribution, or, in finite samples, approximately
\([p (n-1) / (n-p)] F_{p, n-p}\). Hence, in the bivariate case, taking
\(c^2 = \chi^2_2(0.95)= 5.99 \approx 6\) encloses approximately 95\% of
the data points under normal theory. A 68\% coverage data ellipse with
\(c^2 = \chi^2_2(0.68) = 2.28\) gives a bivariate analog of the standard
\(\bar{x} \pm 1 s_x\) and \(\bar{y} \pm 1 s_y\) intervals. See Friendly
et al. (2013) for properties of data ellipsoids and their use to
interpret a wide variety of problems and applications in multivariate
linear models.

In practice, \(p\)-dimensional data ellipsoids can be viewed in variable
space via 2D or 3D projections, or for all \(p\) variables, in a
pairwise scatterplot matrix of 2D projections. Alternatively, they can
be viewed in the space of any linear transformation
\(\mat{Y} \mat{T} \mapsto \mat{Y}^\star\), where the principal
components transformation provides useful views in low-D projections
accounting for maximal total variance.

\subsection{Simple example: Iris data}\label{sec:ex-iris}

\begin{figure}[!htb]

{\centering \includegraphics[width=.49\textwidth]{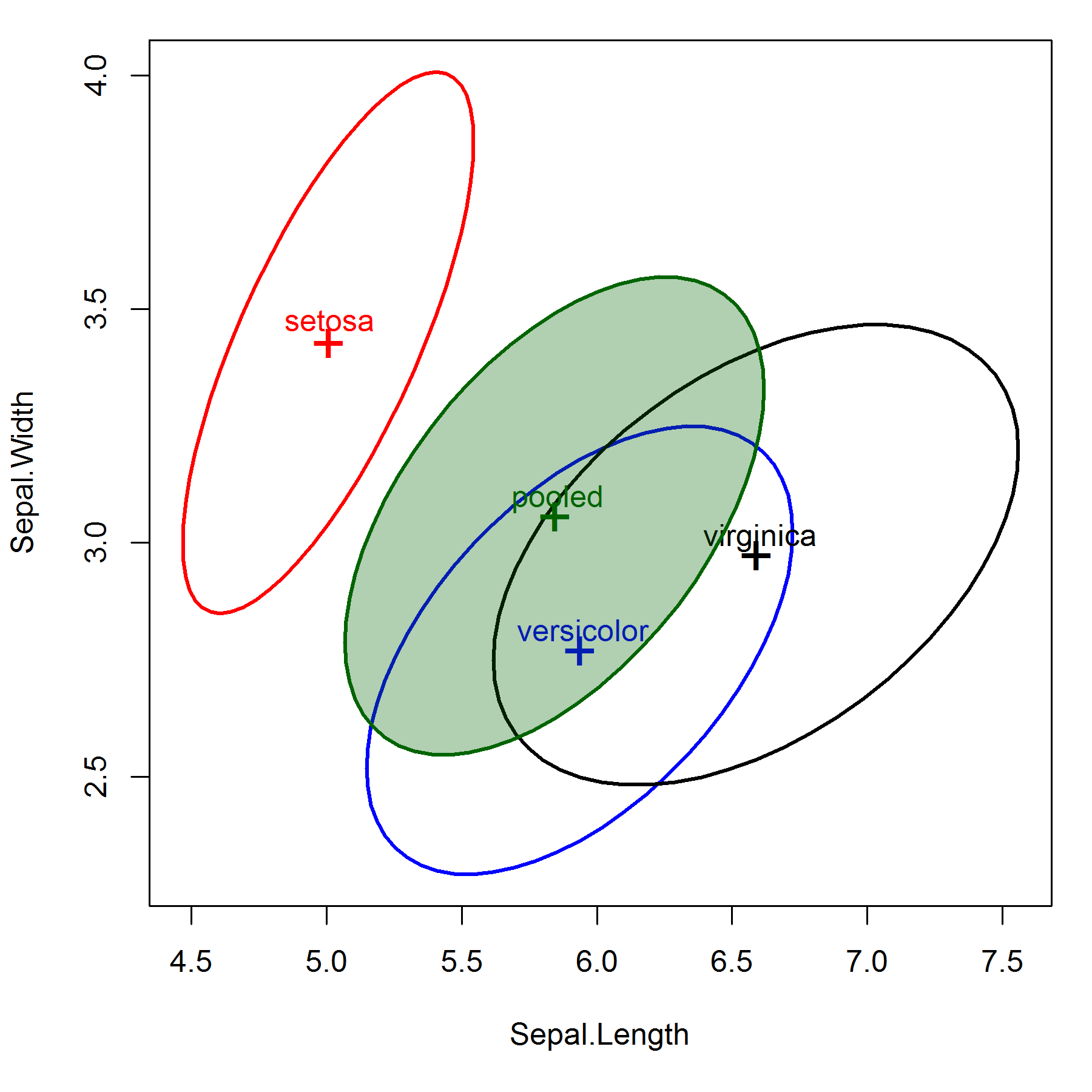} \includegraphics[width=.49\textwidth]{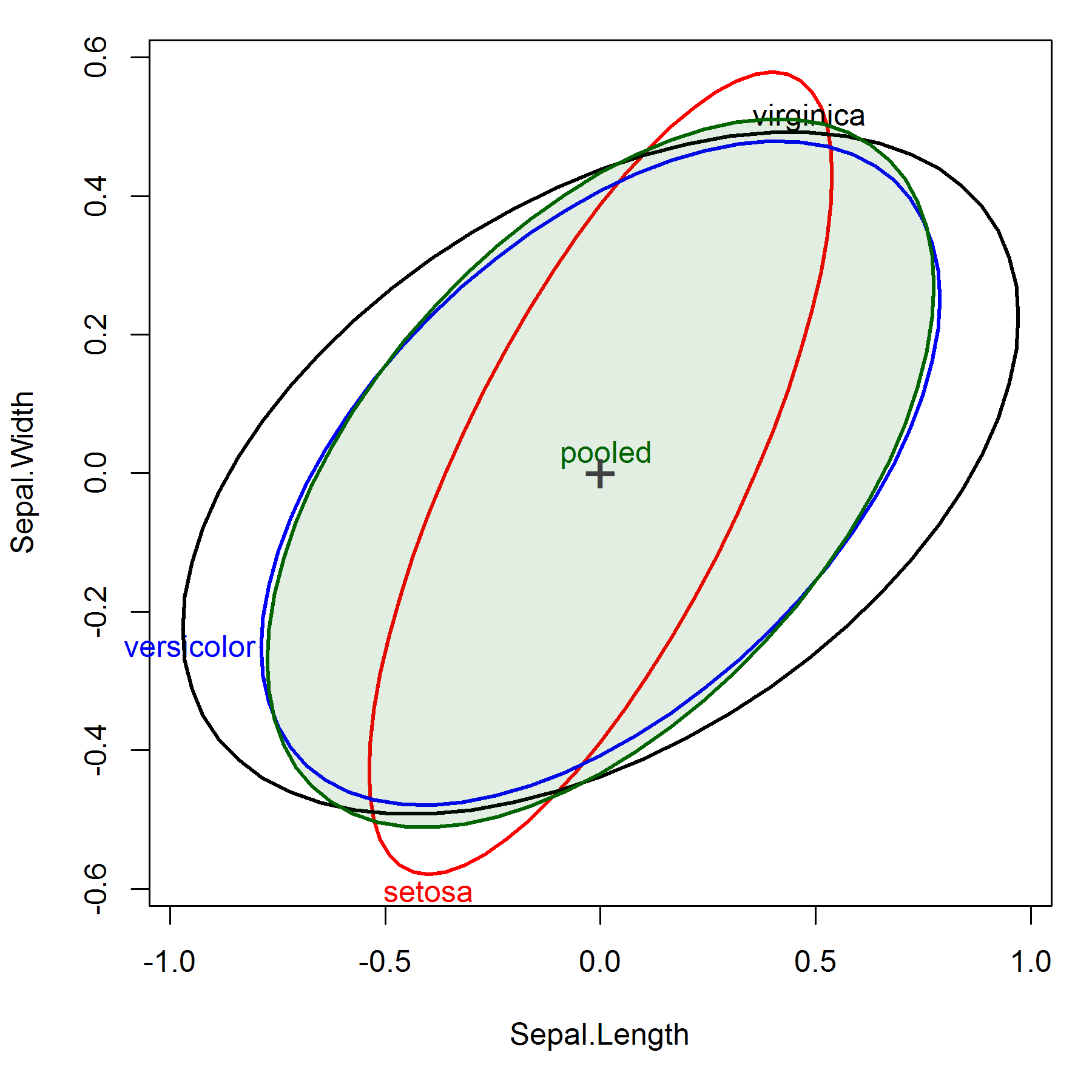} 

}

\caption{Covariance ellipses for the Iris data. Left: separate groups and the pooled within-group covariance matrix; right: all covariance matrices centered at the origin.}\label{fig:iris-covEllipses}
\end{figure}

It is easiest to illustrate these ideas using the well-known Iris data
set (Anderson, 1935), which pertains to four measures (sepal width and
height, and petal width and height) of three species of iris flowers
from the Gaspe Peninsula. One approach to visualizing within group
variability is to begin with an enhanced scatterplot that adds a
standard (68\%) data ellipse for each group. Then imagine taking away
the data points (and other enhancements) leaving only the data ellipses,
and add the corresponding data ellipse for the pooled sample variance
covariance matrix \(\mat{S}_p\). This gives a visual summary of group
means and of the within-group covariance, and is shown in the right
panel of \figref{fig:iris-covEllipses}. In this plot the variances and
covariances look similar for the Versicolor and Virginca groups, but the
Setosa group differs by exhibiting a higher correlation between sepal
length and width and a smaller variance on sepal length.

Finally, we center all the ellipses at the origin in order to focus
\emph{only} on size and shape of the within-group covariances, so that
these can be \emph{directly} compared visually.\footnote{This example
  seems at first glance to be a special case, because all variables are
  measured in the same units. However, the units do not matter in most
  of our plots because the axis ranges are taken from the data and scale
  units are not equated. The plots in \figref{fig:iris-covEllipses}
  would look identical except for tick labels if we transformed sepal
  length from centimeters to inches.} For these two variables, we can
now see that the covariance of \emph{Virginca} is nearly identical to
\(\mat{S}_p\), while \emph{Versicolor} has somewhat greater variance on
sepal length.\footnote{Such plots are produced by the \func{covEllipses}
  function in the \textsf{heplots} package.}

\begin{figure}[!htb]

{\centering \includegraphics[width=.6\textwidth]{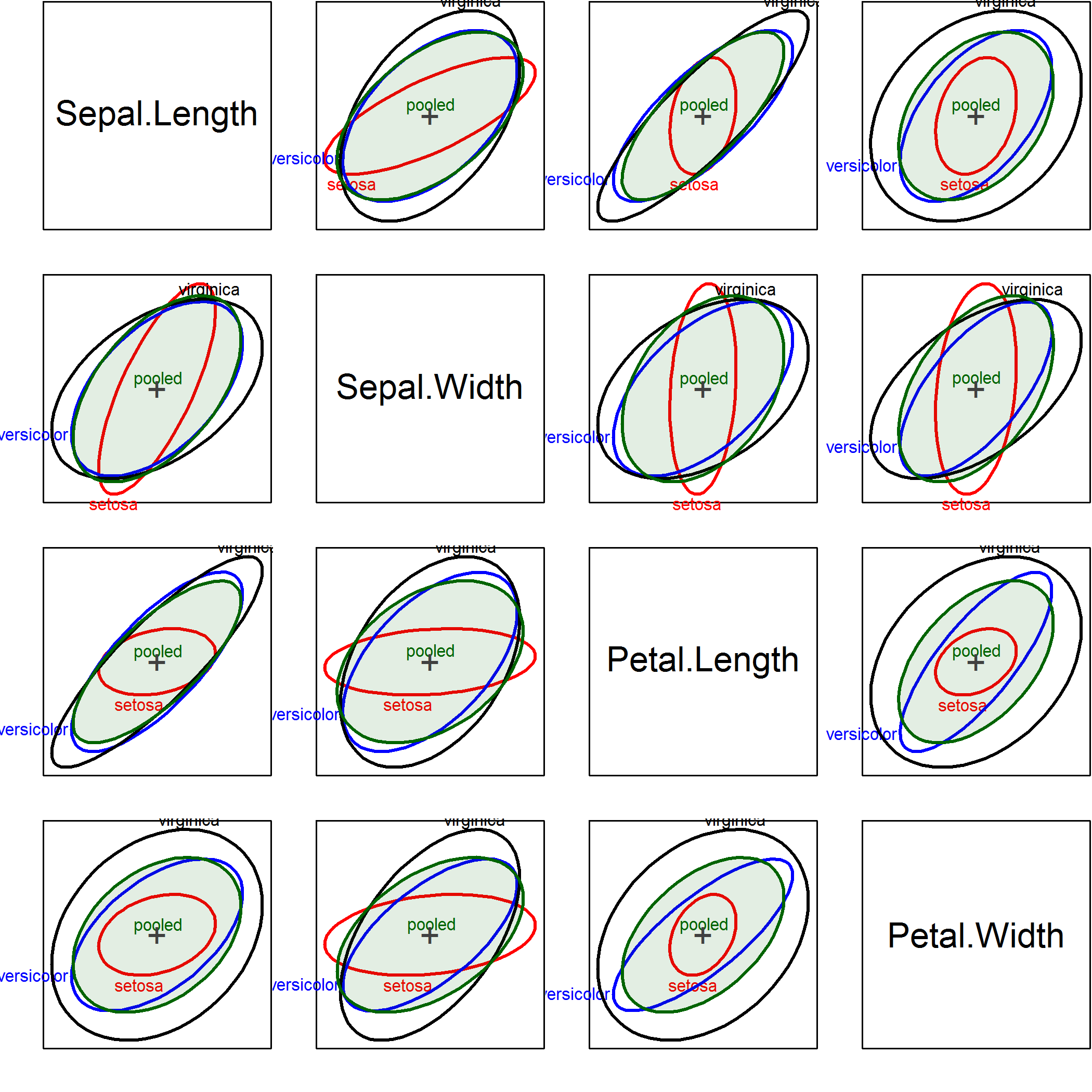} 

}

\caption{Pairwise data ellipses for the Iris data, centered at the origin. This view makes it easy to compare the variances and covariances for all pairs of variables.}\label{fig:iris-cov-pairs2}
\end{figure}

This method becomes particularly useful when we look at the data
ellipses for all pairs of variables in scatterplot matrix format. As in
the right panel of \figref{fig:iris-covEllipses}, we center these
ellipsoids at the origin. The display in \figref{fig:iris-cov-pairs2}
shows only size (variance) and shape (correlation) differences, which
speak directly to the question of homogeneity of covariance matrices.

It can now be seen precisely \emph{how} the covariance matrix for Setosa
differs from those of the other species. The within group correlations
differ for all pairs of variables, and as well, the variances are
noticeably smaller for petal width and petal length. In addition, while
\emph{Versicolor} and \emph{Virginca} have similar shapes, close to that
of the pooled covariance matrix, in most panels (particularly for petal
width), \emph{Virginca} exhibits greater variance.

\subsubsection{More general models}\label{more-general-models}

In these plots, the centered views in \figref{fig:iris-cov-pairs2}
correspond to an analysis of the covariance matrices among the
\emph{residuals} from the MLM predicting the four responses from the
species variable. Consequently, the same ideas apply in more general
models. For example, in a MANCOVA setting, the model may include one or
more quantitative covariates.\footnote{For instance, in R notation,
  \texttt{mod1\ \textless{}-\ lm(cbind(y1,\ y2,\ y3)\ \textasciitilde{}\ Group\ +\ x1\ +\ x2)},
  where three response variables are being predicted by the grouping
  factor and two covariates.} The analyses suggested above could then be
applied to the residuals from this model. Likewise, in a two-way MANOVA
design, with factors \(A\) and \(B\), we could treat the combinations of
these factors as the ``group'' variable and view the pairwise data
ellipses.\footnote{For example, we could estimate such a model within R
  using
  \texttt{mod2\ \textless{}-\ lm(cbind(y1,\ y2,\ y3)\ \textasciitilde{}\ A:B)}
  and then generate the pairwise covariance data ellipses with
  \texttt{covEllipses(residuals(mod2),\ variables=1:3)}.}

\subsection{Low-rank views}\label{low-rank-views}

With \(p>3\) response variables, a simple alternative to the pairwise 2D
projections shown in \figref{fig:iris-cov-pairs2} is the projection into
the principal component space accounting for the greatest amounts of
total variance in the data. For the Iris data, a simple PCA of the
covariance matrix shows that nearly 98\% of total variance in the data
is accounted for in the first two dimensions.

\begin{figure}[!htb]

{\centering \includegraphics[width=\textwidth]{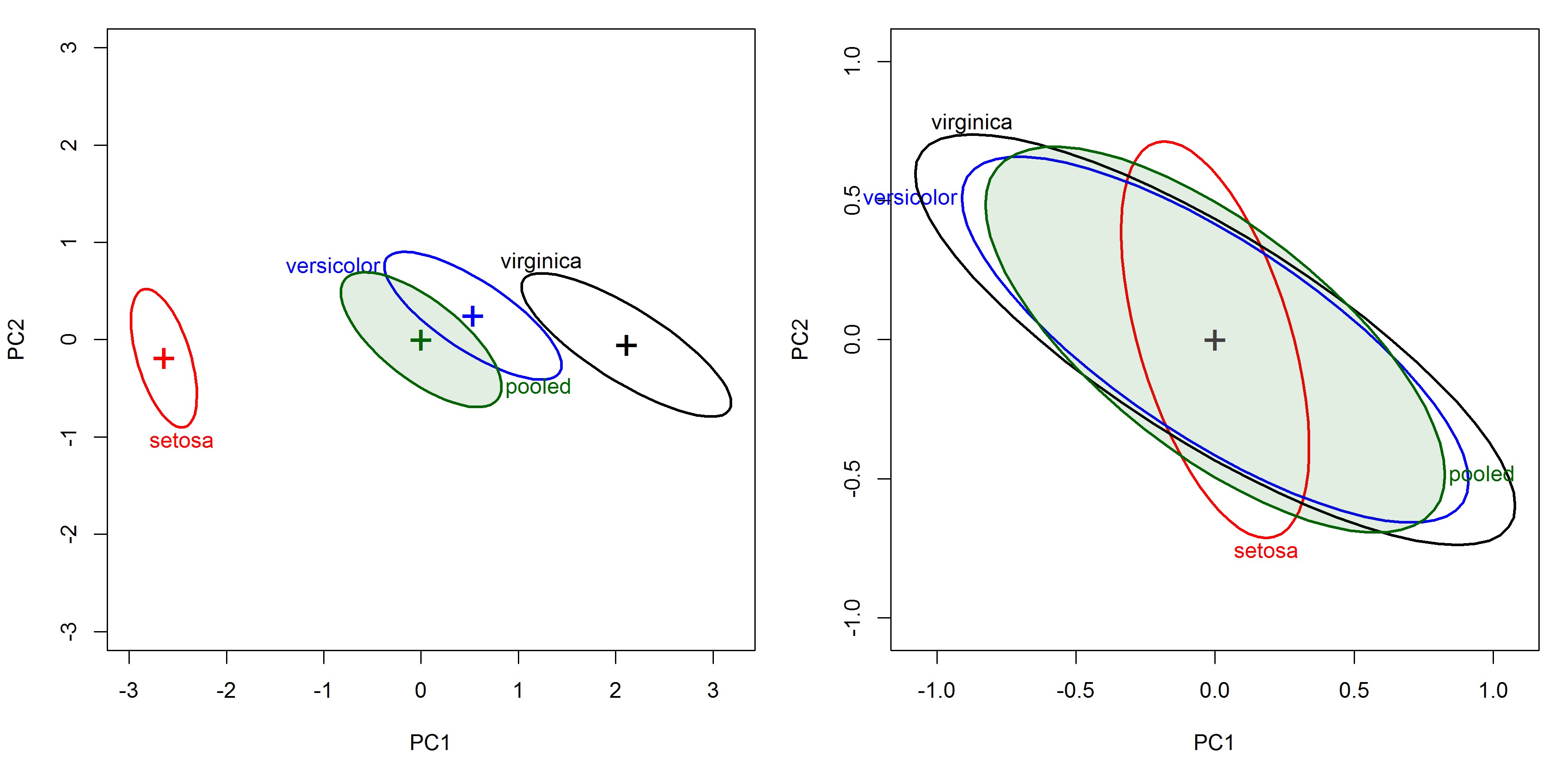} 

}

\caption{Covariance ellipsoids for the first two principal components of the iris data. Left: Uncentered, showing group means on the principal components; right: centered at the origin.}\label{fig:iris-pca}
\end{figure}

\figref{fig:iris-pca} shows the plots of the covariance ellipsoids for
the first two principal component scores, uncentered (left panel) and
centered (right panel). The dominant PC1 (92\% of total variance)
essentially orders the species by a measure of overall size of their
sepals and petals. In the centered view, it can again be seen how
\emph{Setosa} differs in covariance from the other two species, and that
while \emph{Virginca} and \emph{Versicolor} both have similar shapes to
the pooled covariance matrix, Versicolor has somewhat greater variance
on PC1.

We note that PCA is focused on a low-rank approximation to account for
\emph{total variance} of the data. In the MANOVA context, the main
question concerns between-group variance (differences among means)
relative to within-group variance. For this question, views in canonical
space provide the same advantages, as described in the Supplemental
Appendix.

\subsubsection{Small dimensions can
matter}\label{small-dimensions-can-matter}

For the Iris data, the first two principal components account for 98\%
of total variance, so we might think we are done here. Yet, it turns out
that in a variety of multivariate contexts small dimensions can matter.
For example, Friendly \& Kwan (2009) showed that problems of
multicollinearity in regression models could be readily viewed as near
singularities that exist in the space of the \emph{smallest} principal
component dimensions, but cannot be seen in the larger dimensions.
Similarly, multivariate outliers often do not appear in bivariate views
of the data in variable space, but can stand out like sore thumbs in the
space of the smallest PCA dimensions.\footnote{A simple yet powerful
  demonstration: Generate triples, \((x_1, x_2, x_3)\) as \(U[0,1]\) and
  scale each set to unit sum, so all points lie on the simplex
  \(x_1 + x_2 + x_3=1\). Then add a few outliers within a unit sphere of
  radius \(r \le 0.05\) centered at the origin. The outliers will not
  stand out in any univariate or bivariate plots along the coordinate
  axes, but will be dramatic when viewed along the third principal
  component, which is orthogonal to the plane of the simplex.}

As we will see, Box's M test, because it is a (linear) function of all
the eigenvalues of the between and within group covariance matrices, is
also subject to the influence of the smaller dimensions, where
differences among \(\mat{S}_i\) and of \(\mat{S}_p\) can lurk.

\begin{figure}[!htb]

{\centering \includegraphics[width=.5\textwidth]{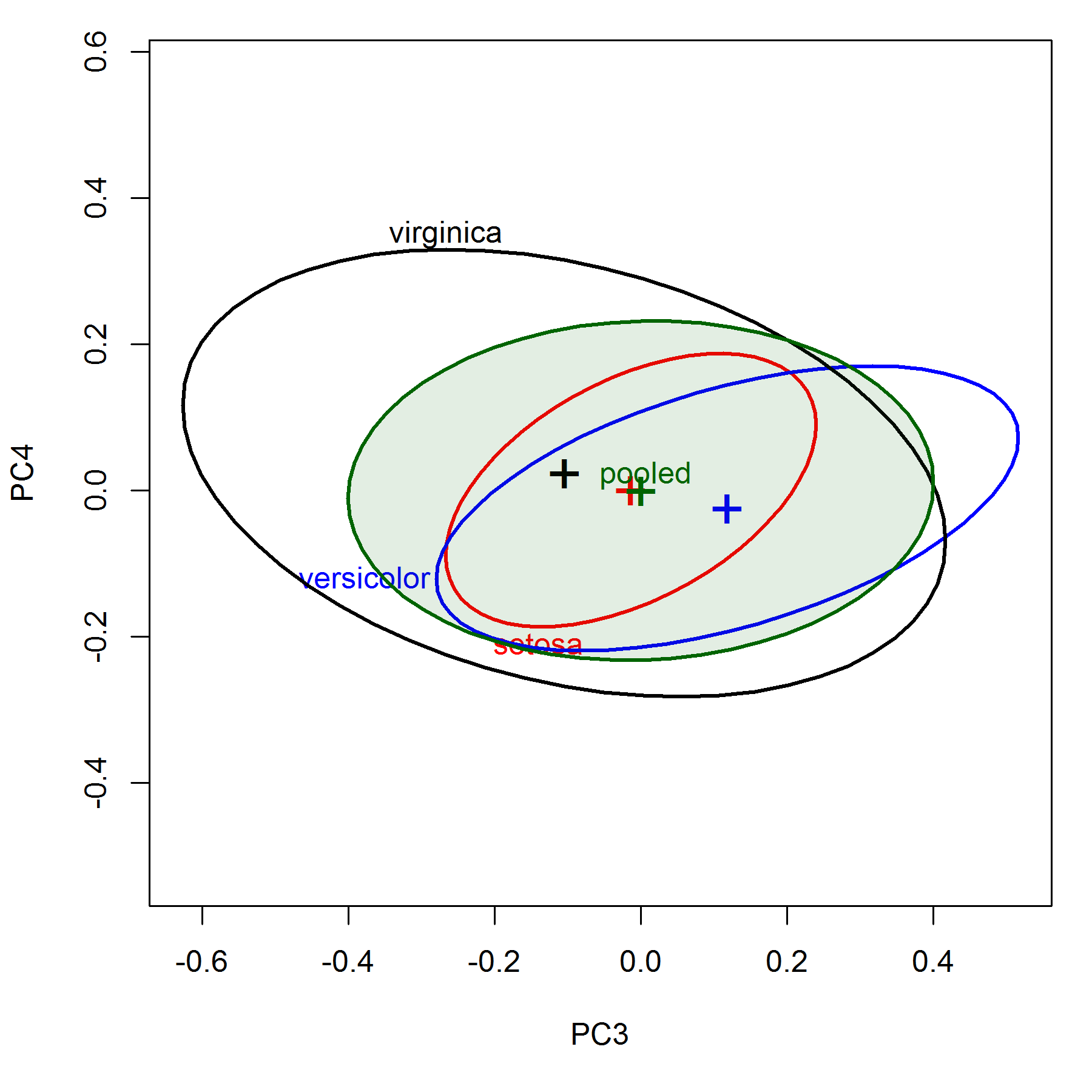} 

}

\caption{Covariance ellipsoids for the last two principal components.}\label{fig:pca-iris-dim34}
\end{figure}

\figref{fig:pca-iris-dim34} shows the covariance ellipsoids in (PC3,
PC4) space. PC3 contrasts \texttt{Sepal.Length} against the other
variables; PC4 contrasts \texttt{Sepal.Length} and \texttt{Petal.Width}
vs. \texttt{Sepal.Width} and \texttt{Petal.Length}. The main point is
that. even though these dimensions contribute little to total variance,
there are more pronounced differences in the within-group shapes
(correlations) relative to the pooled covariance, and these contribute
to a rejection of homogeneity by Box's M test. Here we see that the
correlation for Virginca is of opposite sign from the other two groups.
(The total sample covariance, ignoring Species, is of course
uncorrelated in all principal component dimensions.)

\section{Other examples}\label{other-examples}

In what follows, it will be instructive to use two other empirical
examples to illustrate our graphical methods: (a) one where it turns out
that there are important differences among group means but little
evidence for heterogeneity of covariances; (b) another where there are
both differences in means and heterogeneity, but the number of response
variables is large, which makes understanding these effects more
difficult.

\subsection{Skulls data}\label{sec:ex-skulls}

For comparison with what we have seen for the Iris data, the
\texttt{Skulls} data set provides an example where there are also
substantial differences among the means of groups, but little evidence
for heterogeneity of their covariance matrices.

\begin{figure}[!htb]
\begin{center}
    \includegraphics[width=.6\textwidth]{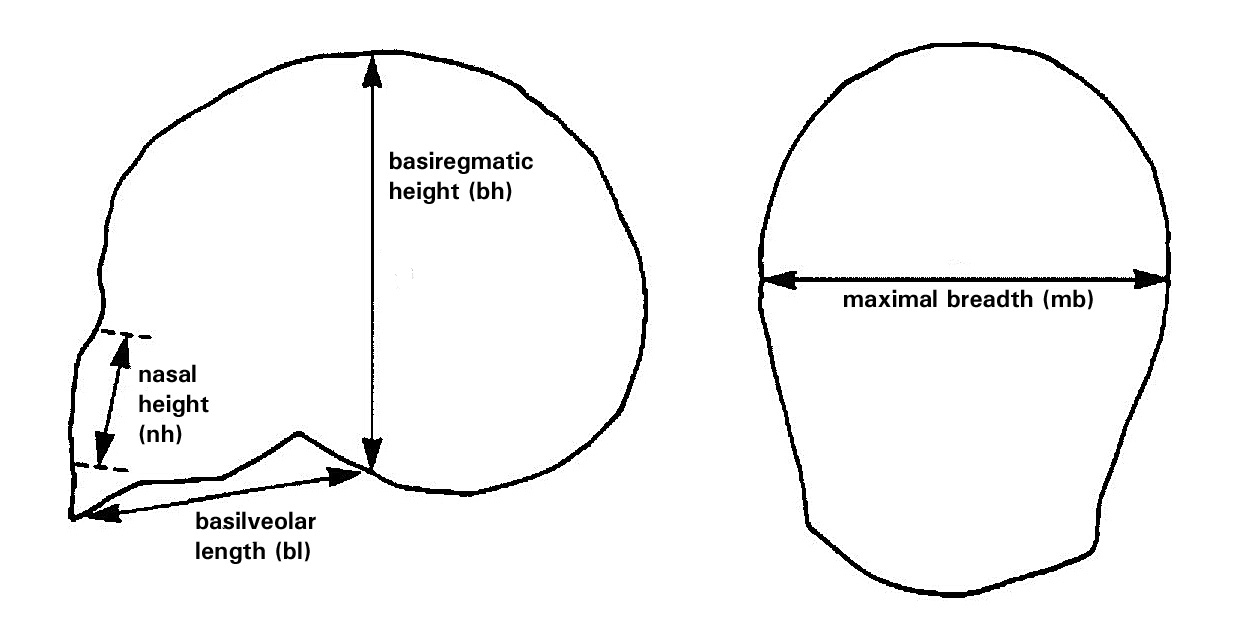}
\caption{Diagram of the skull measurements for the Egyptian skulls data set. Maximal breadth and basibregmatic height are the basic measures of ``size'' of a skull.  Basialveolar length and nasal height are important anthropometric measures of skull ``shape''.}
\label{fig:skulls}
\end{center}
\end{figure}

The data concern four physical measurements of size and shape made on
150 Egyptian skulls from five epochs ranging from 4000 BC to 150 AD. The
measures are: maximal breadth (\texttt{mb}), basibregmatic height
(\texttt{bh}), basialveolar length (\texttt{bl}), and nasal height
(\texttt{nh}) of each skull. See
\url{http://www.redwoods.edu/instruct/agarwin/anth_6_measurements.htm}
for the formal definitions of these measures, and \figref{fig:skulls}
for a diagram of what they pertain to. The question of interest in this
analysis is whether and how these measurements changed over time.
Systematic changes over time in means and/or covariances is of interest
because it could indicate interbreeding among migrant populations (or
the influence of other factors, such as diet).

A MANOVA of this data set shows a highly significant effect of the
\texttt{epoch} factor (Pillai trace = 0.3533, approx.
\(F (16, 434.45)= 3.512, p < 0.000001\)).\footnote{This can be conducted
  as a MLM in \R as follows:
  \texttt{lm(cbind(mb,\ bh,\ bl,\ nh)\ \textasciitilde{}\ epoch,\ data=Skulls)}.}
Treating \texttt{epoch} as an ordered factor yields an even strong test
for linear trend in the means over time, and all non-linear trends are
effectively null. The conclusion so far is that for these measures of
skull size and shape, there are approximately systematic changes over
time.

\begin{figure}[!htb]

{\centering \includegraphics[width=.7\textwidth]{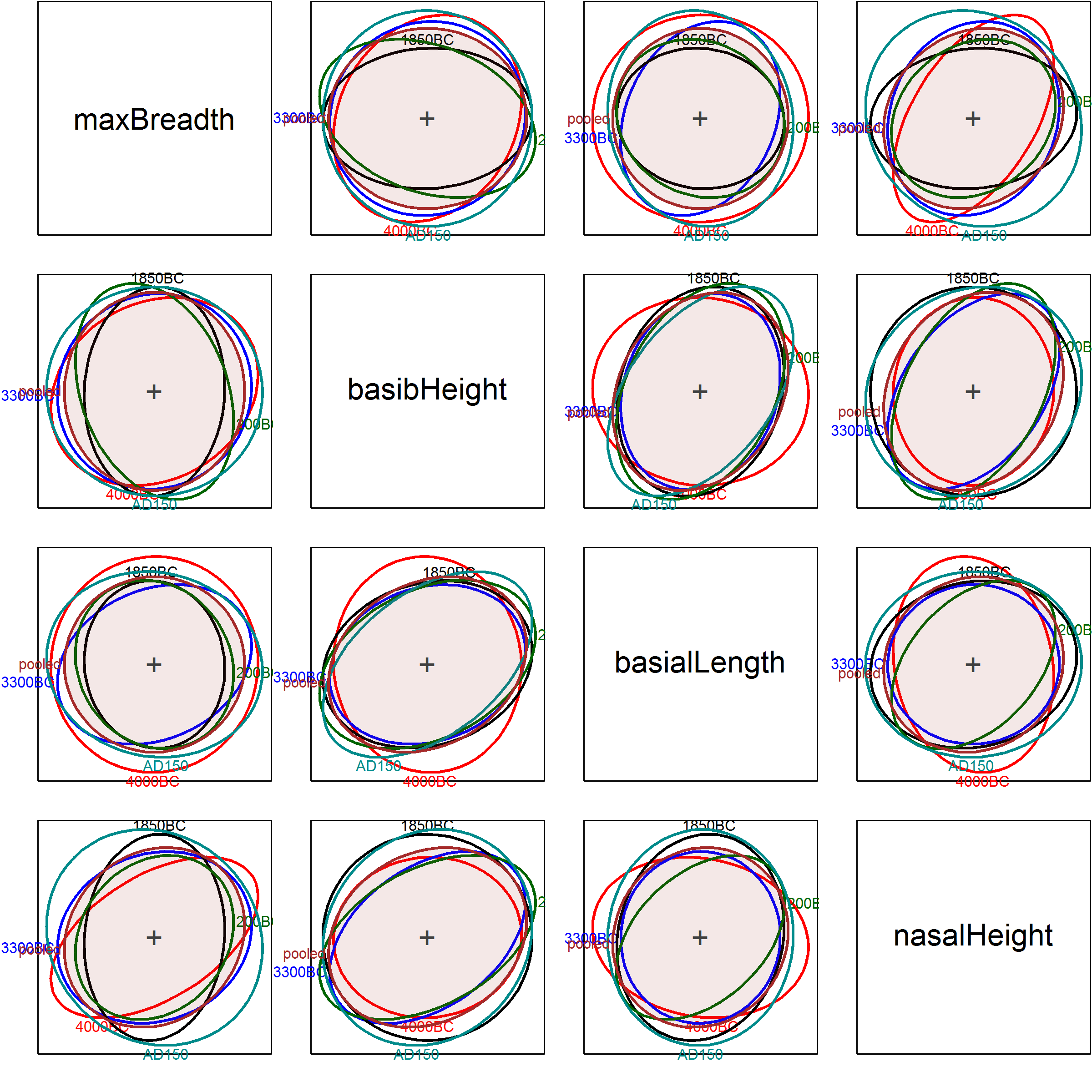} 

}

\caption{Pairwise data ellipses for the Skulls data, centered at the grand mean. Those for the pooled data are shaded.}\label{fig:skulls-cov}
\end{figure}

\figref{fig:skulls-cov} shows the centered covariance ellipsoids for all
epochs and for the pooled data.\footnote{Such figures can be generated
  using the \texttt{covEllipses()} function from the \texttt{heplots}
  package.} For the most part, these are all coincident, indicating
equal covariance matrices. Only for the variable basialveolar length
does any epoch differ perceptibly, where it has slightly greater
variance in the earliest epoch (4000BC). It can also be seen that these
four measures are relatively uncorrelated within each epoch.

\subsection{Wine data}\label{sec:ex-wine}

The \texttt{Wine} data\footnote{This data set is contained in the
  \texttt{candisc} package, and is originally from the UCI Machine
  Learning Repository
  (\url{http://archive.ics.uci.edu/ml/datasets/Wine}).} is a classic in
the machine learning literature as a high-D classification problem, but
is also of interest for examples of MANOVA and discriminant analysis.
These data are the results of a chemical analysis of wines grown in the
same region in Italy but derived from three different cultivars of
grapes: Barolo, Grignolino, and Barbera. The analysis determined the
quantities of 13 constituents found in each of the three types of wines.
The total sample size is \(N = 178\), but the data are unbalanced
(\(n_i = 59, 71, 48\)).

\begin{figure}[!htb]

{\centering \includegraphics{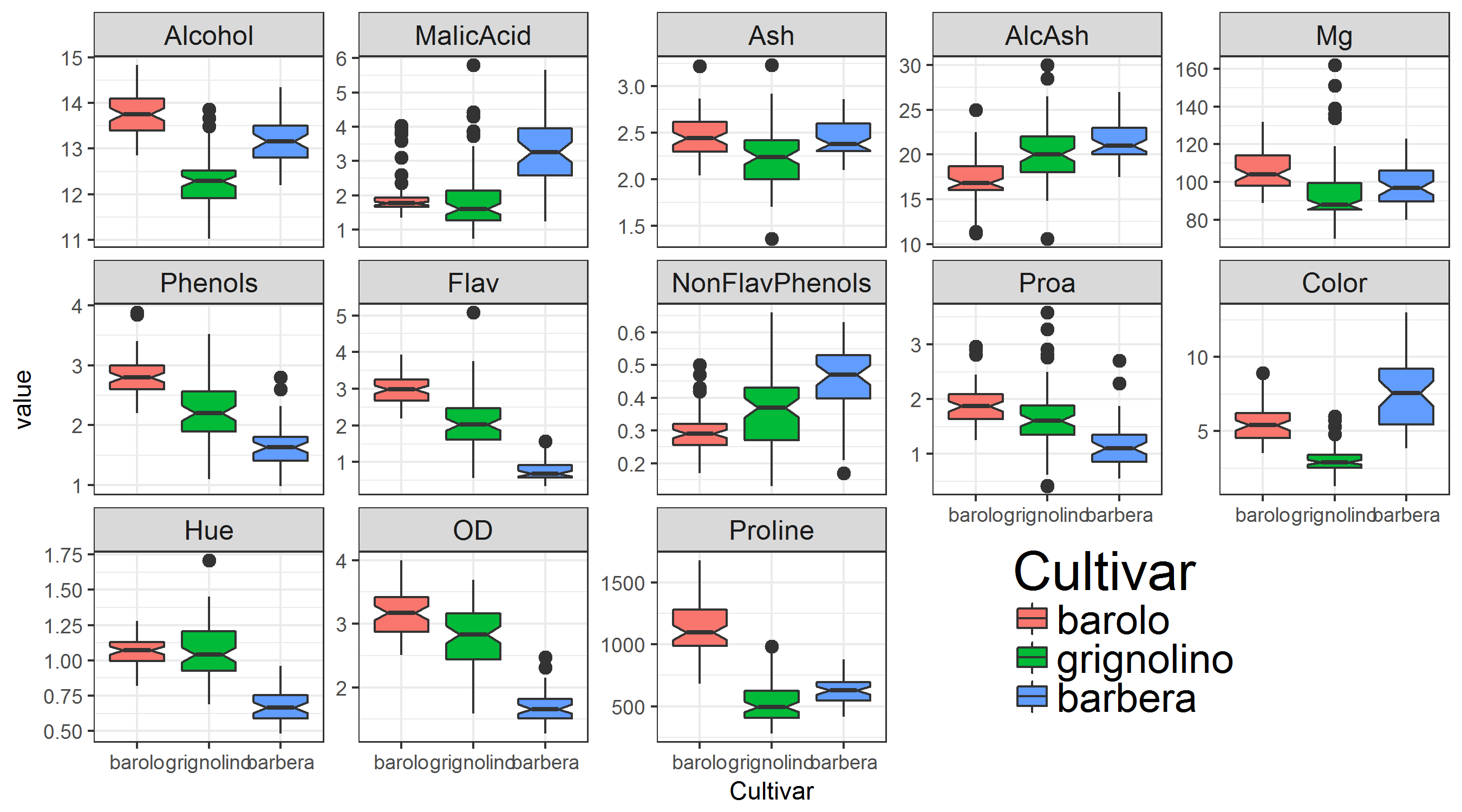} 

}

\caption{Boxplots of the distributions of the Wine variables, by Cultivar. How do the means differ?  How do the variances differ?}\label{fig:wine-boxplots}
\end{figure}

By way of introduction to this data set, the set of boxplots in
\figref{fig:wine-boxplots} for all of the response variables by
\texttt{Cultivar} gives a useful overview. It is easy to see that most
of the variables differ substantially among the cultivars, but the
pattern of differences in means or medians is complex across the
variables. There is also a substantial number of outliers for some of
the variables, particularly \texttt{MalicAcid} and \texttt{Proa}. It is
much harder to characterize how the wines differ in variance, though
differences on some variables appear pronounced (e.g.,
\texttt{MalicAcid}, \texttt{Flav}, \texttt{Color}).

\section{Visualizating Box's M test}\label{sec:vis-boxm}

The covariance ellipse plots we have seen in earlier examples (e.g.,
\figref{fig:iris-covEllipses} and \figref{fig:skulls-cov}) are useful
schematic summaries, but in some cases, a simpler visual summary might
be more useful. \eqref{eq:boxm} suggests that the simplest visualization
might focus on the components of Box's M test, for example, a dot plot
of the log determinants of the covariance matrices \(\mat{S}_i\)
together with that of the pooled \(\mat{S}_p\). To the extent that the
covariance matrices are all equal, so too should the values on which
Box's test are based.

An important virtue of these plots is that they can show \emph{how} the
groups differ from each other, and from the pooled covariance matrix on
the scalar measure \(\ln \vert \mathbf{S} \vert\). In this way, they can
suggest more specific questions or hypotheses regarding the equality of
covariance matrices, analogous to the use of contrasts and linear
hypotheses for testing differences among group mean vectors.

Such plots are far more useful with \textbf{confidence intervals} around
the \(\ln \vert \mathbf{S}_i \vert\) and
\(\ln \vert \mathbf{S}_p \vert\). Recently, Cai, Liang, \& Zhou (2015)
have suggested an asymptotic, central-limit theorem approximation to the
distribution of \(\log \vert \mathbf{S} \vert\).\footnote{Box's M test
  is calculated by the function \func{boxM} in \texttt{heplots}. These
  plots are produced by the \func{plot} method for \class{boxM} objects.}

\begin{figure}[!htb]

{\centering \includegraphics[width=\textwidth]{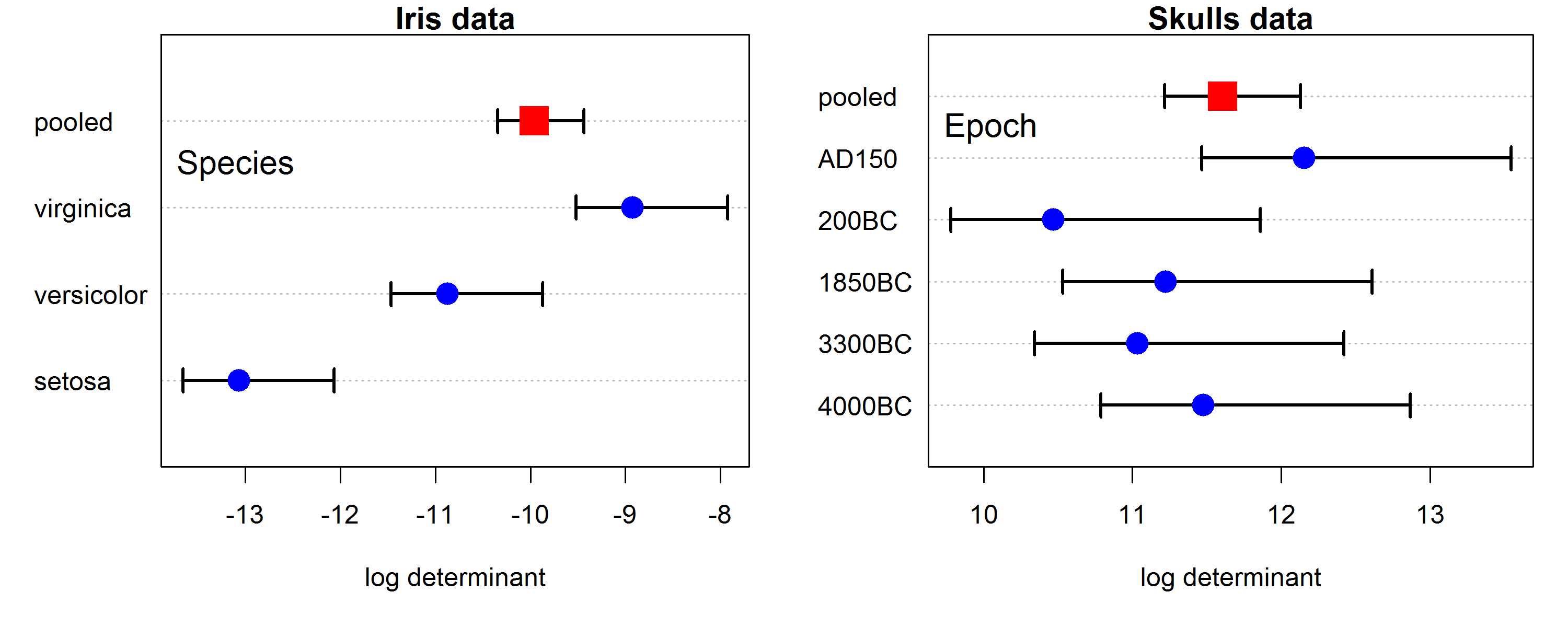} 

}

\caption{Plots of log determinants of the components of Box's M test with asymptotic 95\% confidence intervals. Left: The Iris data shows substantial heterogeneity; right: the Skulls data shows some small differences, but no evidence for heterogeneity.}\label{fig:boxm-plots}
\end{figure}

To illustrate, Box's M test gives an approximate chi-square,
\(\chi^2 (20) = 140.94, p < 2.2 \times 10^{-16}\) for the Iris data,
while the Skulls data gives \(\chi^2 (40) = 45.67, p = 0.248\). The
correponding plots for these tests are shown in \figref{fig:boxm-plots}.
For the Iris data (left), \emph{Setosa} stands out having a
substantially smaller covariance matrix (by
\(\log \vert \mathbf{S} \vert\)) than the other species. The intervals
for \emph{Versicolor} and \emph{Virginica} overlap with that for
\(\ln \vert \mathbf{S}_p \vert\), but seem to differ from each other.

In contrast, for the \texttt{Skulls} data (\secref{sec:ex-skulls}), the
plot of the log determinants in \figref{fig:boxm-plots} (right) shows
that the 95\% confidence intervals for the \(\ln \vert \mat{S}_i \vert\)
all overlap with each other and with that for the pooled
\(\ln \vert \mat{S}_p \vert\).

Although these differences among covariance matrices are not significant
by Box's M test, we can use this example to illustrate how such plots
can suggest scientifically meaningful hypotheses regarding the equality
of covariance matrices, analogous to what we are accustomed to doing
with tests for mean differences. For the sake of an example, assume that
changes in variances and covariances of such skull measurements is of
interest, and that we were able to obtain a sample 10 times as large
from each epoch, giving the same pattern of results, but with standard
errors divided by \(\sqrt{10}\). We might then try to interpret the
general decrease in \(\ln \vert \mat{S}_i \vert\) from the earliest
epoch to 200BC. Were skulls becoming more homogeneous over time? But,
what happened in the 150 AD sample?

\begin{figure}[!htb]

{\centering \includegraphics[width=.7\textwidth]{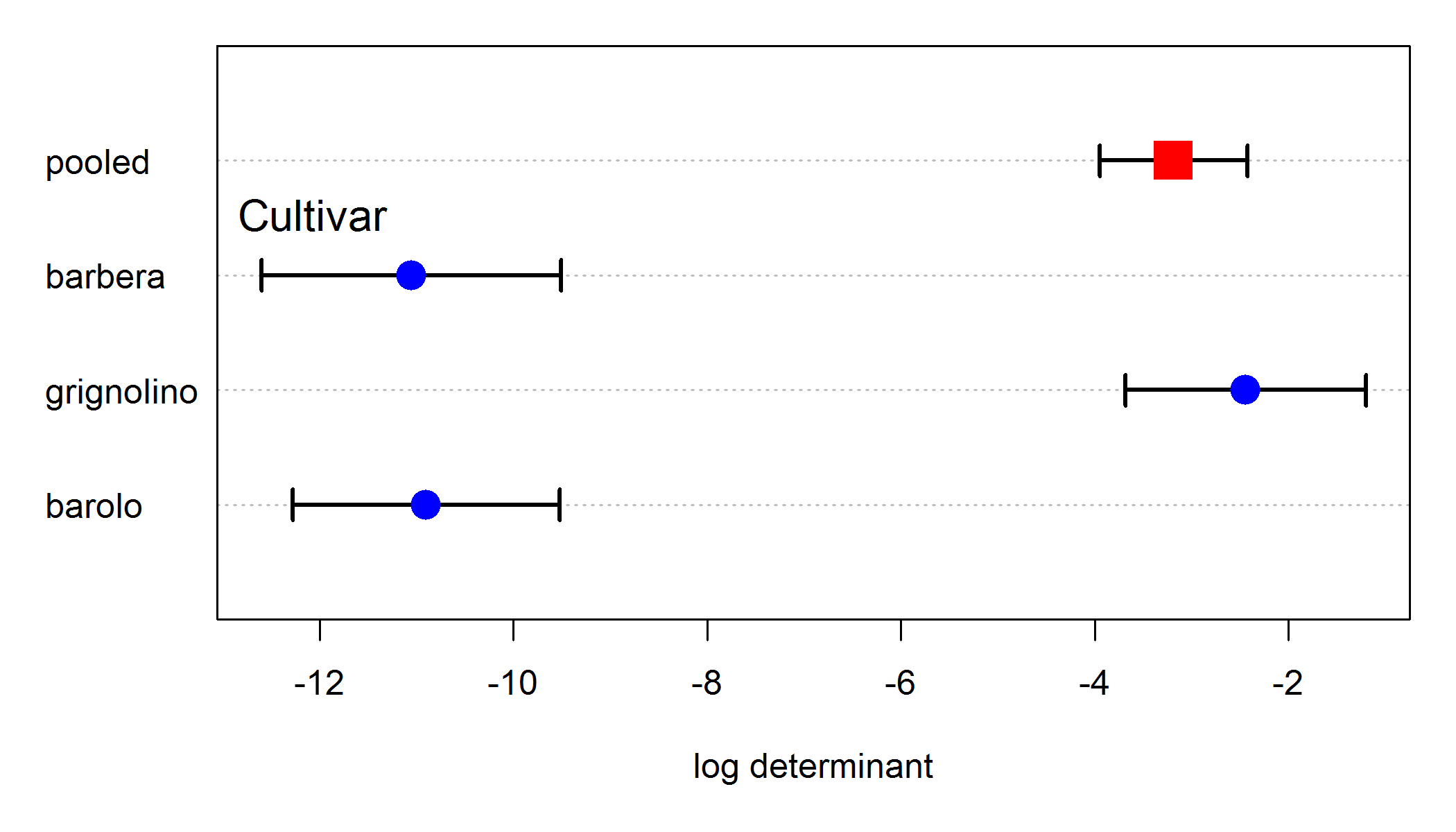} 

}

\caption{Plot of log determinants of the covariance matrices for the Wine data.}\label{fig:wine-boxm-plot}
\end{figure}

For completeness, and use below, \figref{fig:wine-boxm-plot} shows the
same type of plot for the Wine data. Box's M test is overwhelmingly
significant, \(\chi^2 (182) = 684.2, p < 2.2 \times 10^{-16}\), and the
figure shows why: the \(\ln \vert \mat{S}_i \vert\) for Barbera and
Barolo differ substantially from that of the pooled
\(\ln \vert \mat{S}_p \vert\) and from that of Grignolino, which is
closest to that of the pooled covariance matrix.\\
But the Wine data has 13 response variables; the Supplementary Appendix
describes some other graphical methods that help to understand this in a
2D canonical space.

\subsection{Eigenvalue plots}\label{sec:eigval-plots}

Having made some progress with visualizing the components of Box's M
test, it is natural to ask if other plots or other test statistics can
address these relationships in a more nuanced manner. In the MLM, the
various test statistics (Wilks' \(\Lambda\), Hotelling-Lawley and Pillai
Trace criteria, Roy's maximum root test) are all functions of the
eigenvalues of a hypothesis matrix \(\mat{H}\) relative to an error
matrix \(\mat{E}\). So too, all reasonable test statistics for equality
of covariance matrices are functions of the eigenvalues of the
\(\mathbf{S}_i\) and \(\mathbf{S}_p\).

Another sensible plot is therefore an analog of a \emph{scree plot} of
eigenvalue versus dimension number, with separate curves for each
matrix, similar to their use in exploratory factor analysis.

\begin{figure}[!htb]

{\centering \includegraphics{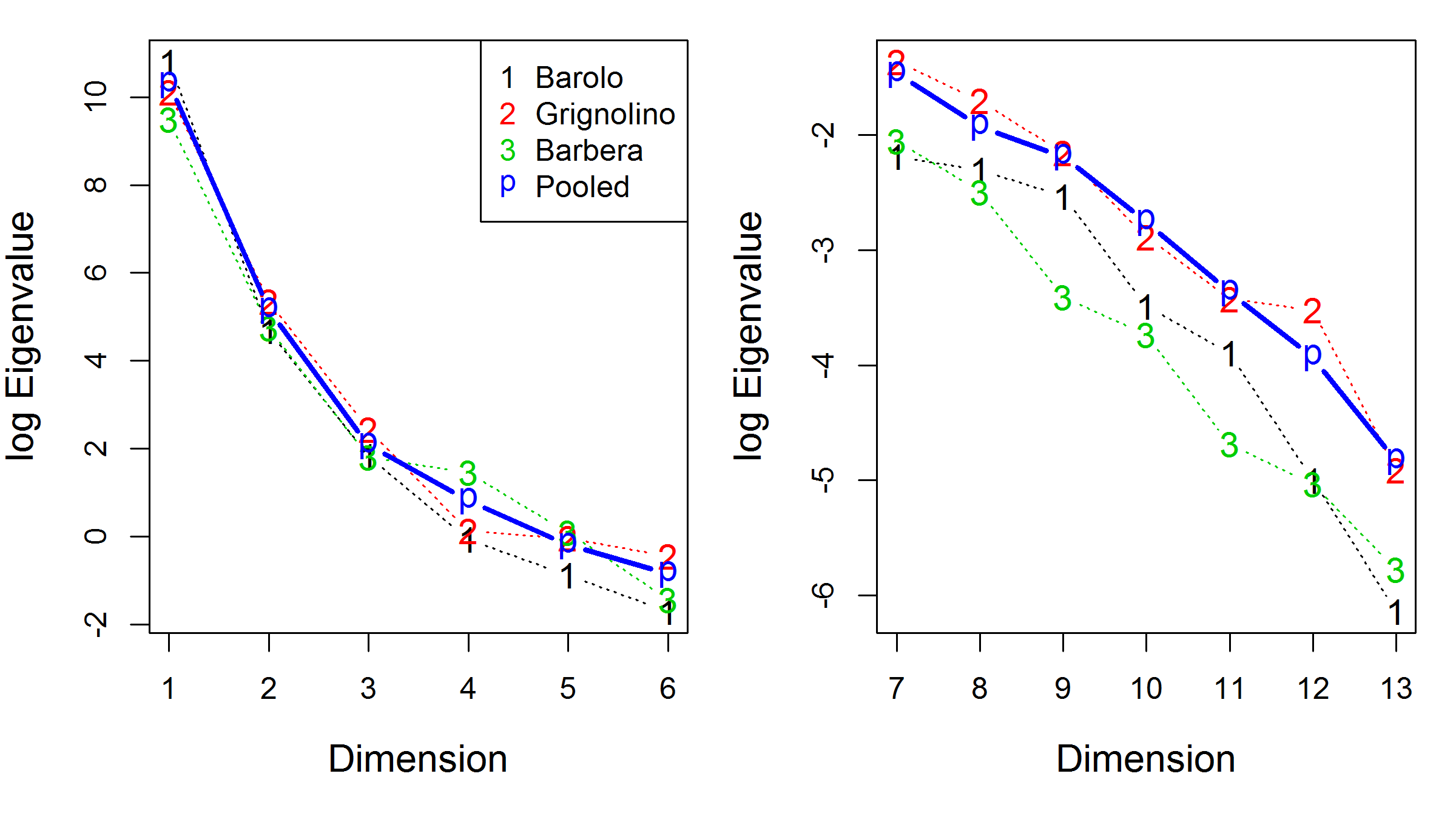} 

}

\caption{Scree plots of log eigenvalues of the covariance matrices for the Wine data. Those for the pooled covariance matrix are shown with a heavier line, marked `p'.}\label{fig:wine-matplots}
\end{figure}

\figref{fig:wine-matplots} shows such a plot for the Wine data. This
dataset is comprised of 13 response variables, so Box's test is based on
13 eigenvalues. To preserve resolution, we show eigenvalues on the log
scale, and in two separate panels.

It can be seen that the eigenvalues on the largest dimensions do not
differ very much across the groups. However, they differ progressively
more among the groups on the dimensions with small eigenvalues. The
differences among the groups in the right panel of
\figref{fig:wine-matplots} are similar to what was seen in the the log
determinants plot (\figref{fig:wine-boxm-plot}): Grignolino (group 2) is
quite close to the result for the pooled covariance matrix, while the
Barolo (group 1) and Barbera (group 3) wines differ. This demonstrates
that Box's M test is indeed sensitive to differences among the smaller
eigenvalues.

\subsection{Other test statistics}\label{sec:other-stats}

As we saw above (\secref{sec:vis-cov}), the question of equality of
covariance matrices can be expressed in terms of the similarity in size
and shape of the data ellipses for the individual group \(\mat{S}_i\)
relative to that of \(\mat{S}_p\). Box's M test uses just one possible
function to describe this size: the logs of their determinants.

When \(\mat{\Sigma}\) is the covariance matrix of a multivariate vector
\(\vec{y}\) with eigenvalues
\(\lambda_1 \ge \lambda_2 \ge \dots \lambda_p\), the properties shown in
\tabref{tab:ell-size} represent methods of describing the size and shape
of the ellipsoid in \(\Real{p}\). More general theory and statistical
applications of the geometry of ellispoids is given by Friendly et al.
(2013).

\begin{table}[!htb]
\caption{Statistical and geometrical properties of ``size'' of an ellipsoid}\label{tab:ell-size}
\begin{tabular}{llll}
    Size                   &  Conceptual formula                    & Geometry       & Function \\
\hline
(a) Generalized variance:  & $\det{\mat{\Sigma}} = \prod_i \lambda_i$ & area, (hyper)volume & geometric mean\\  
(b) Average variance:        & $\trace{\mat{\Sigma}} = \sum_i \lambda_i $ & linear sum & arithmetic mean\\     %% superscripts turned to subscripts / GM
(c) Average precision:        & $1/ \trace{\mat{\Sigma}^{-1}} = 1/\sum_i (1/\lambda_i) $ &  & harmonic mean\\
(d) Maximal variance:      & $\lambda_1$ & maximum dimension & supremum
 \end{tabular}
\end{table}

Hence, for a sample covariance matrix \(\mat{S}\),
\(\vert \mat{S} \vert\) is a measure of generalized variance and
\(\ln \vert \mat{S} \vert\) is a measure of average variance across the
\(p\) dimensions.

\begin{figure}[!htb]

{\centering \includegraphics{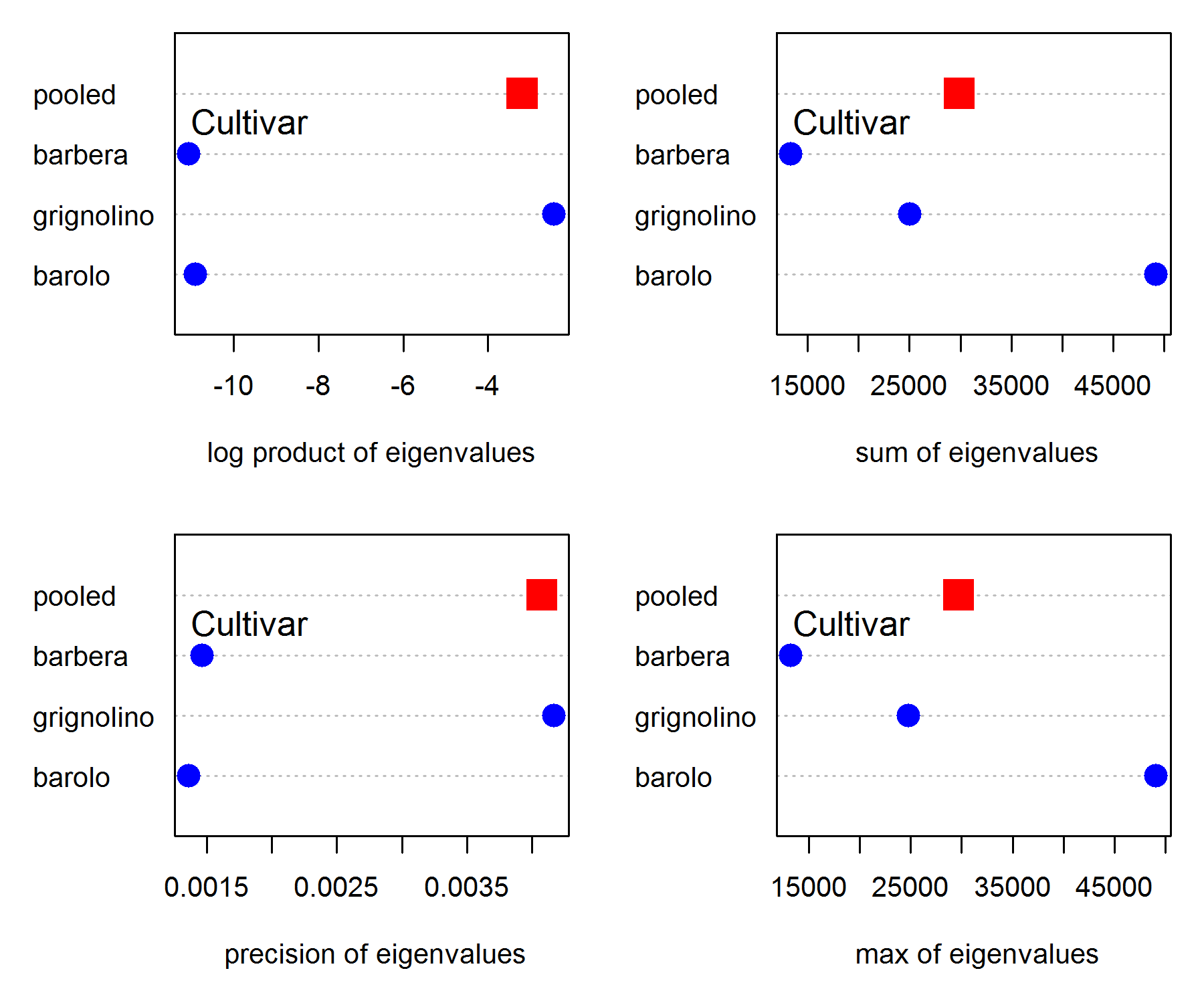} 

}

\caption{Plot of eigenvalue statistics of the covariance matrices for the Wine data.}\label{fig:wine-others}
\end{figure}

The \class{boxM} methods in \pkg{heplots} can compute and plot all of
the functions of the eigenvalues in \tabref{tab:ell-size}. The results
are shown in \figref{fig:wine-others}.

Except for the absence of error bars, the plot for log product in
\figref{fig:wine-others} (upper left panel) is the same as that in
\figref{fig:wine-boxm-plot}. In principle, it is possible to add such
confidence intervals for all these measures through the use of
bootstrapping, but this has not yet been implemented.

For this data set, the pattern of points in the plot for Box's M is also
more or less the same as that for the precision measure. The plots for
the sum of and maximum eigenvalue are also similar to each other, but
differ from those of the two measures in the left column of
\figref{fig:wine-others}. The main point is that these are not \emph{all
the same}, so different functions reflect different patterns of the
eigenvalues.\footnote{In analogous examples for other data sets we also
  see different patterns over measures, but these are not the same as in
  this example.}

The sum of eigenvalues is similar in form to the Hotelling-Lawley trace
criterion used in MLM test and the maximum eigenvalue is similar to
Roy's test, and all these test statistics have convenient \(F\)
approximations. Plots such as \figref{fig:wine-others} and those we have
tried for other examples suggest that it would be useful to develop
analogous tests for homogeneity of covariance matrices, with possibly
different properties of error rates and power against specified
alternatives. We do not pursue this topic here, but leave it open for
further research.

\section{Concluding Remarks}\label{concluding-remarks}

The idea for this paper arose from discussion in a graduate course on
multivariate data analysis in connection with the topic of the
assumptions of MANOVA (independence, multivariate normality of
residuals, homogeneity of covariance matrices), diagnostic tests for
these, and remedies when they are violated. The instructor (the first
author) had presented Box's M test, with the caveat of the opening
quotation by George Box that the test was not robust against
mild-to-moderate non-normality, while the Pillai-Bartlett trace test for
mean differences does possess this robustness.

One student commented that this was akin to recommending a screening
test for lung cancer that was more sensitive to influenza. Another
student asked, ``Well, if I run Box's test and it shows significance,
how can I decide if my MANOVA results are valid? How can I determine
which groups differ in covariance matrices and on which variables?''

These seemed to be perfectly reasonable questions, and while we were
aware of the modifications that might account for non-normality
(O'Brien, 1992; Tiku \& Balakrishnan, 1984), these are not typically
available in standard software and therefore their theory is cold
comfort to applied researchers. More importantly, we were struck with
how little information is provided by the result of such a significance
test, regardless of its Type I error and power properties. What was
lacking in the null hypothesis \(p\)-value was any \emph{insight} into
the nature of group differences in covariance matrices.

The approach we have outlined here stems from the mantras of exploratory
data analysis (EDA) in a multivariate setting:

\begin{quote}
The greatest value of a picture is when it forces us to notice what we
never expected to see (Tukey, 1977, p. vi).
\end{quote}

\begin{quote}
The purpose of {[}data{]} display is comparison (recognition of
phenomena), not numbers (Tukey, 1990).
\end{quote}

We began with the idea of displaying covariance matrices directly in
terms of data ellipsoids that serve as normal theory minimally
sufficient visual summaries. We mentioned, but did not illustrate, that
classical estimates of \(\mat{S}\) could be replaced by robust MVE and
MCD alternatives in the presence of multivariate outliers.

The simple dot plot (\figref{fig:wine-boxm-plot}) of the components of
Box's M test answers the first question: How do the groups differ in
covariance matrices? It does more than this, however, because it
suggests that \emph{other} functions of the eigenvalues of the
\(\mat{S}_i\) and \(\mat{S}_p\) might provide alternative measures of
homogeneity of covariance matrices (\figref{fig:wine-others}), and that
their distribution across the orthogonal dimensions of within-group
variation provides valuable insight into the properties of these
measures and statistical tests.

The HE plot framework we have described provides some answers to the
second question: If the multivariate responses do differ among groups in
their variances and covariances, which responses contribute to this and
how do they differ? We formulated an extension of the univariate
Levene--Brown-Forsythe test to the MANOVA setting, with the property
that HE plots and canonical discriminant HE plots for the question of
mean differences apply directly to the question of homogeneity of
covariance matrices. Pairwise HE plots in variable space show which
responses differ in scatter, and their projection into canonical space
(Figure 2 in the supplementary appendix) provides a convenient 2D
representation, often with quite a simple interpretation.

Finally, we do not fear that this paper might attract the opprobrium
that we ``flung data onto many canvases to see what stuck.'' To the
contrary, we were struck by the relative degree of consistency of the
resulting plots across these various visual approaches. When covariance
ellipsoids showed differences among groups in variable space or
principal components space, we could understand how these differences
were reflected in the Box's M plots. All of this is much more satisfying
than a \(p\)-value, robust or otherwise.

\section{Supplementary Materials}\label{sec:supmat}

A Supplemental Appendix describes the HE plot framework for tests of
mean differences and gives some further examples. It also describes a
multivariate extension of Levene's test within this framework. The
\R packages \textsf{candisc} and \textsf{heplots} are freely available
from the Comprehensive R Archive Network,
\url{http://cran.us.r-project.org/}. Complete \R scripts for the Iris,
Wine, and Skulls examples are available at
\url{https://mattsigal.github.io/eqcov_supp/}.

\section{Acknowledgements}\label{acknowledgements}

We are grateful to Augustine Wong and Harrison Zhou for discussions on
the asymptotic distribution of log determinants of covariance matrices.
Wayne Oldford provided much useful feedback on an initial draft of the
paper. Two anonymous reviewers and the associate editor gave detailed
critical and constructive feedback that helped us greatly to revise the
article.

\section*{References}\label{references}
\addcontentsline{toc}{section}{References}

\hypertarget{refs}{}
\hypertarget{ref-Van_Aelst:2011}{}
Aelst, S. V., \& Willems, G. (2011). Robust and efficient one-way MANOVA
tests. \emph{Journal of the American Statistical Association},
\emph{106}(494), 706--718.
\url{http://doi.org/10.1198/jasa.2011.tm09748}

\hypertarget{ref-Anderson:35}{}
Anderson, E. (1935). The irises of the Gaspé peninsula. \emph{Bulletin
of the American Iris Society}, \emph{35}, 2--5.

\hypertarget{ref-Bartlett:1937}{}
Bartlett, M. S. (1937). Properties of sufficiency and statistical tests.
\emph{Proceedings of the Royal Society of London. Series A},
\emph{160}(901), 268--282. \url{http://doi.org/10.2307/96803}

\hypertarget{ref-Box:1949}{}
Box, G. E. P. (1949). A general distribution theory for a class of
likelihood criteria. \emph{Biometrika}, \emph{36}(3-4), 317--346.
\url{http://doi.org/10.1093/biomet/36.3-4.317}

\hypertarget{ref-Box:1950}{}
Box, G. E. P. (1950). Problems in the analysis of growth and wear
curves. \emph{Biometrics}, \emph{6}, 362--389.

\hypertarget{ref-Box:1953}{}
Box, G. E. P. (1953). Non-normality and tests on variances.
\emph{Biometrika}, \emph{40}(3/4), 318--335.
\url{http://doi.org/10.2307/2333350}

\hypertarget{ref-BrownForsythe:1974}{}
Brown, M. B., \& Forsythe, A. B. (1974). Robust tests for equality of
variances. \emph{Journal of the American Statistical Association},
\emph{69}(346), 364--367.
\url{http://doi.org/10.1080/01621459.1974.10482955}

\hypertarget{ref-Cai-etal:2015}{}
Cai, T. T., Liang, T., \& Zhou, H. H. (2015). Law of log determinant of
sample covariance matrix and optimal estimation of differential entropy
for high-dimensional gaussian distributions. \emph{Journal of
Multivariate Analysis}, \emph{137}, 161--172.
\url{http://doi.org/https://doi.org/10.1016/j.jmva.2015.02.003}

\hypertarget{ref-Cochran:1941}{}
Cochran, W. G. (1941). The distribution of the largest of a set of
estimated variances as a fraction of their total. \emph{Annals of
Eugenics}, \emph{11}(1), 47--52.
\url{http://doi.org/10.1111/j.1469-1809.1941.tb02271.x}

\hypertarget{ref-Conover-etal:1981}{}
Conover, W. J., Johnson, M. E., \& Johnson, M. M. (1981). A comparative
study of tests for homogeneity of variances, with applications to the
outer continental shelf bidding data. \emph{Technometrics},
\emph{23}(4), 351--361.
\url{http://doi.org/10.1080/00401706.1981.10487680}

\hypertarget{ref-Dempster:69}{}
Dempster, A. P. (1969). \emph{Elements of continuous multivariate
analysis}. Reading, MA: Addison-Wesley.

\hypertarget{ref-Finch:2013}{}
Finch, H., \& French, B. (2013). A monte carlo comparison of robust
MANOVA test statistics. \emph{Journal of Modern Applied Statistical
Methods}, \emph{12}(2), 35--81.
\url{http://doi.org/10.22237/jmasm/1383278580}

\hypertarget{ref-Friendly:07:manova}{}
Friendly, M. (2007). HE plots for multivariate general linear models.
\emph{Journal of Computational and Graphical Statistics}, \emph{16}(2),
421--444. \url{http://doi.org/10.1198/106186007X208407}

\hypertarget{ref-FriendlyKwan:2009}{}
Friendly, M., \& Kwan, E. (2009). Where's Waldo: Visualizing
collinearity diagnostics. \emph{The American Statistician},
\emph{63}(1), 56--65.

\hypertarget{ref-Friendly-etal:ellipses:2013}{}
Friendly, M., Monette, G., \& Fox, J. (2013). Elliptical insights:
Understanding statistical methods through elliptical geometry.
\emph{Statistical Science}, \emph{28}(1), 1--39.
\url{http://doi.org/http://dx.doi.org/10.1214/12-STS402}

\hypertarget{ref-Gastwirth-etal:2009}{}
Gastwirth, J. L., Gel, Y. R., \& Miao, W. (2009). The impact of Levene's
test of equality of variances on statistical theory and practice.
\emph{Statistical Science}, \emph{24}(3), 343--360.
\url{http://doi.org/10.1214/09-STS301}

\hypertarget{ref-Hakstian:1979}{}
Hakstian, A. R., Roed, J. C., \& Lind, J. C. (1979). Two-sample t-2
procedure and the assumption of homogeneous covariance matrices.
\emph{Psychological Bulletin}, \emph{86}(6), 1255--1263.
\url{http://doi.org/10.1037/0033-2909.86.6.1255}

\hypertarget{ref-Hartley:1950}{}
Hartley, H. O. (1950). The use of range in analysis of variance.
\emph{Biometrika}, \emph{37}(3--4), 271--280.
\url{http://doi.org/10.1093/biomet/37.3-4.271}

\hypertarget{ref-Harwell:1992}{}
Harwell, M. R., Rubinstein, E. N., Hayes, W. S., \& Olds, C. C. (1992).
Summarizing monte carlo results in methodological research: The one- and
two-factor fixed effects ANOVA cases. \emph{Journal of Educational and
Behavioral Statistics}, \emph{17}(4), 315--339.
\url{http://doi.org/10.3102/10769986017004315}

\hypertarget{ref-Levene:1960}{}
Levene, H. (1960). Robust tests for equality of variances. In I. Olkin,
S. G. Ghurye, W. Hoeffding, W. G. Madow, \& H. B. Mann (Eds.),
\emph{Contributions to probability and statistics: Essays in honor of
Harold Hotelling} (pp. 278--292). Stanford, Calif: Stanford University
Press.

\hypertarget{ref-Lix:1996}{}
Lix, J., L.M. Keselman, \& Keselman, H. (1996). Consequences of
assumption violations revisited: A quantitative review of alternatives
to the one-way analysis of variance F test. \emph{Review of Educational
Research}, \emph{66}(4), 579--619.
\url{http://doi.org/10.3102/00346543066004579}

\hypertarget{ref-Olson:1974}{}
Olson, C. L. (1974). Comparative robustness of six tests in multivariate
analysis of variance. \emph{Journal of the American Statistical
Association}, \emph{69}(348), 894--908.
\url{http://doi.org/10.1080/01621459.1974.10480224}

\hypertarget{ref-OBrien:1992}{}
O'Brien, P. C. (1992). Robust procedures for testing equality of
covariance matrices. \emph{Biometrics}, \emph{48}(3), 819--827.
Retrieved from \url{http://www.jstor.org/stable/2532347}

\hypertarget{ref-Rcore:2015}{}
R Core Team. (2015). \emph{R: A language and environment for statistical
computing}. Vienna, Austria: R Foundation for Statistical Computing.
Retrieved from \url{http://www.R-project.org/}

\hypertarget{ref-Rogan:1977}{}
Rogan, J. C., \& Keselman, H. J. (1977). Is the ANOVA f-test robust to
variance heterogeneity when sample sizes are equal?: An investigation
via a coefficient of variation. \emph{American Educational Research
Journal}, \emph{14}(4), 493--498.
\url{http://doi.org/10.3102/00028312014004493}

\hypertarget{ref-TikuBalakrishnan:1984}{}
Tiku, M., \& Balakrishnan, N. (1984). Testing equality of population
variances the robust way. \emph{Communications in Statistics - Theory
and Methods}, \emph{13}(17), 2143--2159.
\url{http://doi.org/10.1080/03610928408828818}

\hypertarget{ref-Timm:75}{}
Timm, N. H. (1975). \emph{Multivariate analysis with applications in
education and psychology}. Belmont, CA: Wadsworth (Brooks/Cole).

\hypertarget{ref-Todorov:2010}{}
Todorov, V., \& Filzmoser, P. (2010). Robust statistic for the one-way
MANOVA. \emph{Computational Statistics \& Data Analysis}, \emph{54}(1),
37--48. \url{http://doi.org/10.1016/j.csda.2009.08.015}

\hypertarget{ref-Tukey:77}{}
Tukey, J. W. (1977). \emph{Exploratory data analysis}. Reading, MA:
Addison Wesley.

\hypertarget{ref-Tukey:90}{}
Tukey, J. W. (1990). Data-based graphics: Visual display in the decades
to come. \emph{Statistical Science}, \emph{5}(3), 327--339.

\hypertarget{ref-Welch:1947}{}
Welch, B. L. (1947). The generalization of ``student's'' problem when
several different population varlances are involved. \emph{Biometrika},
\emph{34}(1--2), 28--35. \url{http://doi.org/10.1093/biomet/34.1-2.28}

\hypertarget{ref-ZhangBoos:1992:BCV}{}
Zhang, J., \& Boos, D. D. (1992). Bootstrap critical values for testing
homogeneity of covariance matrices. \emph{Journal of the American
Statistical Association}, \emph{87}(418), 425--429. Retrieved from
\url{http://www.jstor.org/stable/2290273}

\end{document}